\documentclass[aps, showpacs, pra,amssymb, amsmath, twocolumn]{revtex4}
\usepackage{amsmath,latexsym,amssymb}
\usepackage{graphicx}
\usepackage{amsmath}
\usepackage{latexsym}
\usepackage{amssymb}
\usepackage{bm}
\usepackage{placeins}
\usepackage{color}
\usepackage{bbm}
\usepackage{multirow}
\usepackage{cancel}

\begin{document}

\title{Direct frequency-comb spectroscopy of $6S_{1/2}$-$8S_{1/2}$ transitions of atomic cesium}
\author{Kyung-tae Kim and Jaewook Ahn}
\address{Department of Physics, KAIST, Daejeon 305-701, Korea}
\date{\today}

\begin{abstract}
Direct frequency-comb spectroscopy is used to probe the absolute frequencies of $6S_{1/2}$-$8S_{1/2}$ two-photon transitions of atomic cesium in hot vapor environment. By utilizing the coherent control method of temporally splitting the laser spectrum above and below the two-photon resonance frequency, Doppler-free absorption is built in two spatially distinct locations and imaged for high-precision spectroscopy. Theoretical analysis finds that these transition lines are measured with uncertainty below $5\times10^{-10}$, mainly contributed from laser-induced AC Stark shift.
\end{abstract}

\maketitle

\section{Introduction}
\noindent
Frequency-comb laser generates a train of equally time-separated optical pulses and its spectrum is a comb of equally spaced frequency components, given by 
\begin{equation}
f_n=f_{\rm ceo}+n f_{\rm rep},
\end{equation} 
where $f_{\rm ceo}$ is the carrier-envelope offset frequency, often locked to a radio-frequency standard such as atomic clock, $f_{\rm ref}$ is the 
the comb tooth spacing, which is the repetition frequency of a mode-locked laser, and $n$ is an integer on the order of million~\cite{UdemNature2002}. This laser provides absolute frequencies in optical frequency domain that can be fine-tuned electronically. This laser can provide super-narrow response, in particular, to two-photon absorption, when being used in a counter-propagating beam geometry so that the sum of the frequencies of absorbed frequency comb modes coincides with the resonance. This Doppler-free spectroscopy scheme was initially proposed for the $1S$-$2S$ transition of hydroden~\cite{BaklanovAP1977} and later experimentally demonstrated~\cite{PartheyPRL2011}. This scheme has been widely used for various precision measurements in fundamental constants~\cite{MohrRMP2008, FischerPRL2004}, molecules and ions~\cite{DickensonPRL2013, MeyerPRL2000,ResenbandScience2008}, and even distance-ranging applications~\cite{CoddingtonNatPhotonics2009}.
 
Recently a coherent control method, which rather directly uses the frequency comb for spectroscopy than referencing continuous-wave (CW) lasers is developed~\cite{BarmesNP2013}. Termed as direct frequency-comb spectroscopy (DFCS), this method extends the usage of the frequency-comb even to Doppler-free spectroscopy of atoms in hot vapor environments. For example, rubidium $5S$ to $7S$ transition lines were measured with an enhanced accuracy~\cite{BarmesPRL2013}. This method is relatively simple to experimentally implement, compared to cold-atom based spectroscopy and thus reduces systematic effects such as radiation pressure in cold atom ensembles~\cite{MarianScience2004}.

In this experiment, we probe the $6S_{1/2}$ ($F=3,4$) $\rightarrow$ $8S_{1/2}$ ($F'=3,4$) two-photon transitions of atomic cesium ($^{133}$Cs). As shown in Fig.~\ref{fig1}(a), these transitions have the excitation frequencies around $2\times 365$~THz (822/2~nm in wavelength).
As to be described below, we used the direct frequency-comb spectroscopy method adopted from Ref.~\cite{BarmesPRL2013} and the result is compared with the previous measurement performed with picosecond lasers frequency-stabilized to frequency comb references~\cite{FendelOL2007}.

\section{Measurement principle and setup}

\begin{figure*}[htb]
\centering
\includegraphics[width=0.90\textwidth]{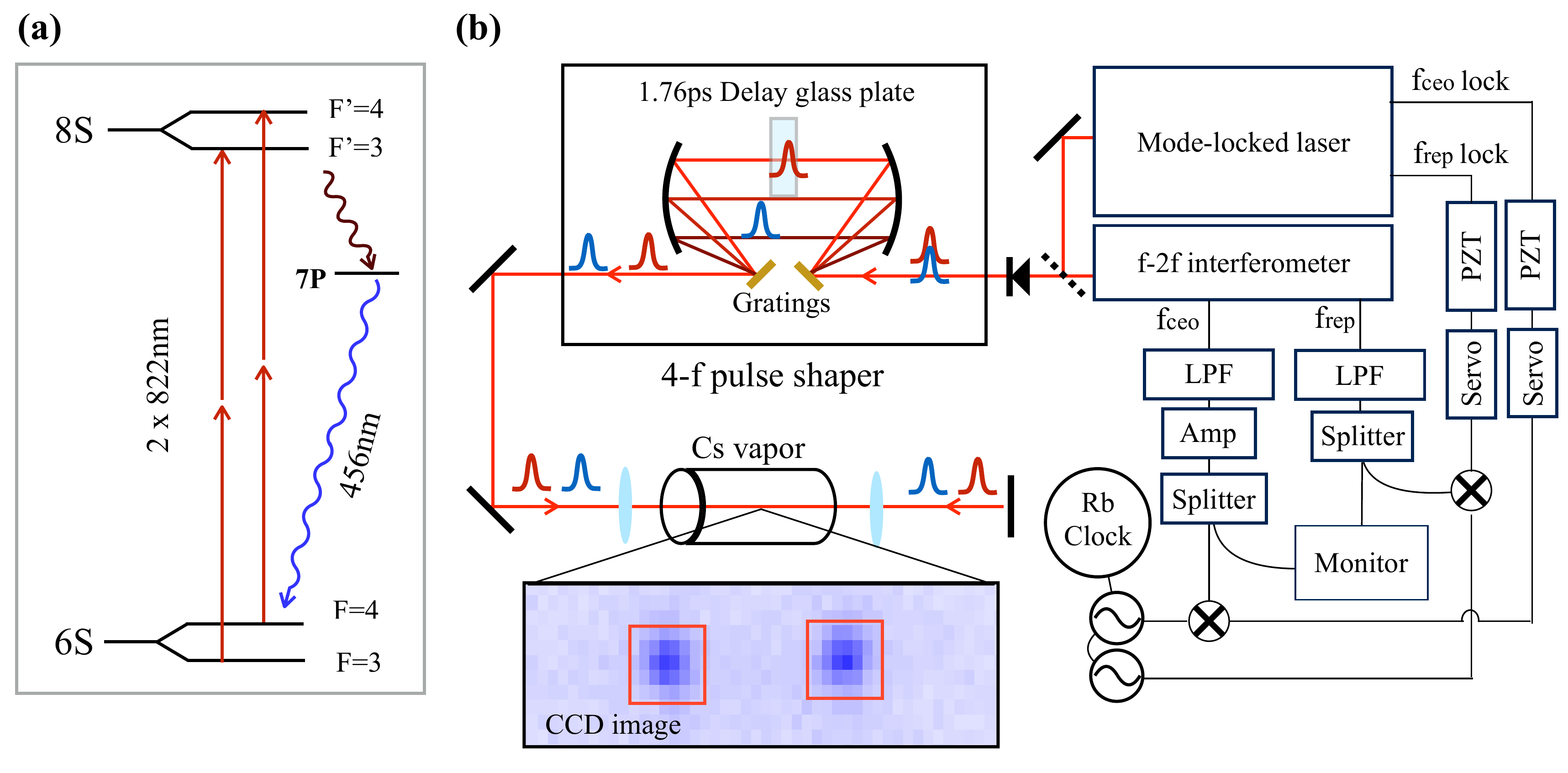}
\caption{(color online) (a) Energy-level structure of atomic cesium ($^{133}$Cs). The two-photon transitions from $6S_{1/2} (F=3,4)$ to $8S_{1/2} (F'=3,4)$, having a two-photon resonance at around 822~nm. After the excitation, the 456~nm fluorescence from $7P$ was collected with a charge-coupled device (CCD) camera to record the population excited to the $8S_{1/2}$ levels. (b) Experimental setup of DFCS. (PZT: piezoelectric inducer, LPF: low pass filter, Amp: amplifier)}
\label{fig1}
\end{figure*}

Experimental setup for our direct frequency-comb spectroscopy is schematically illustrated in Fig.~\ref{fig1}(b). We used a home-made Kerr-lens mode-locked Ti:sapphire laser-oscillator which produced laser pulses frequency-centered at $f_L=365$~THz ($\lambda_L=822$~nm in wavelength) with a bandwidth of $\Delta f=$15~THz (FWHM, $\Delta\lambda=35$~nm). The pulse repetition-rate was controlled in the range from $f_{\rm rep}=80$ to 90~MHz and the carrier-envelope offset frequency from $f_{\rm ceo}=10$ to $30$~MHz. Scanning the frequency within the comb spacing was performed with two comb-parameters, $f_{\rm rep}$ and $f_{\rm ceo}$, that were fine-tuned with the tilt of one cavity mirror (the high reflector) and the displacement of the other (the output coupler) in combination. Scanning resolution was less than 1\%, limited by our detector resolution. The laser was then frequency-stabilized with the conventional $f$-to-$2f$ self-referencing Mach-Zehnder interferometer~\cite{JonesScience2000}, where both $f_{\rm rep}$ and $f_{\rm ceo}$ were locked to a rubidium atomic-clock using custom-made phase-locked feedback loops.

Doppler-broadening of the two-photon absorption from the broad-band single-pulse was avoided with a coherent control method~\cite{BarmesPRL2013}. The initial pulse spectrum was divided into two with respect to the exact two-photon center (the frequency that corresponded to the half of the $6S_{1/2}$-$8S_{1/2}$ transition frequency). When we denote red (blue) pulse for the spectrum below (above) the two-photon center, the red and blue pulses should be applied to the atom at the same time, to satisfy the energy conservation of the two-photon transition. The red and blue pulses were separated in time and  we created a replica of these two pulses that propagates in the opposite direction. Then, at two distinct positions, each red and blue pulse collided with its back-propagating counterpart. In this case, because the direction of the red (blue) and its counterpart were opposite with each other, Doppler-free two-photon absorption (except residual shift due to the frequency difference between the red and blue) occurred at these two positions [see the CCD image on Fig.~\ref{fig1}(b)]. 

However, because the lifetime of the excited state ($\approx1$~$\mu$s) is longer that the typical pulse-repetition time ($\approx 10$~ns), the excited state population can be coherently accumulated as the pulse train passes by. The amount of the accumulated population depends on the delay ($1/f_{\rm rep}$) and the phase difference ($\propto f_{\rm ceo}$) between subsequent pulse pairs. It can be also understood in the frequency domain: As the delay and the phase difference are the two comb parameters, $f_{\rm ceo}$ and $f_{\rm rep}$, the situation becomes simply the summation of two-photon transition from the pairs of two CW lasers whose frequencies are determined by the comb parameters. This explanation works for the weak-field regime that corresponds to our experimental condition. 

In the setup shown in Fig.~\ref{fig1}(b), each laser pulse was split to a pair of sub-pulses with a conventional 4-$f$ geometry pulse-shaper~\cite{WoojunLeePRA2015}. By placing a glass plate on the Fourier plane, the red spectrum pulse was time-delayed by 1.76~ps, with respect to the blue part. No significant pulse broadening (less than 30~fs FWHM) ensured that this plate gave no significant higher-order dispersion. We also controlled the laser power by cutting the laser spectrum with a knife edge on the Fourier plane. The red and blue sub-pulses counter-propagated and were focused with a beam waist of 50~$\mu$m before they interacted with the atoms at the center of the vapor cell. Doppler-free $8S_{1/2}$ excitation occurred at two distinct spatial locations. We measured the $8S_{1/2}$ state population through $8S\rightarrow 7P \rightarrow 6S$ decay channel, where the 456~nm fluorescence ($7P \rightarrow 6S$) was imaged with a charge-coupled device (CCD). The typical image is shown in the Fig.~\ref{fig1}(b), and we integrated the red boxes at the image to extract spectroscopy signal. 

\section{Result and discussion}

\begin{figure}[t]
\centering
\includegraphics[width=0.5\textwidth]{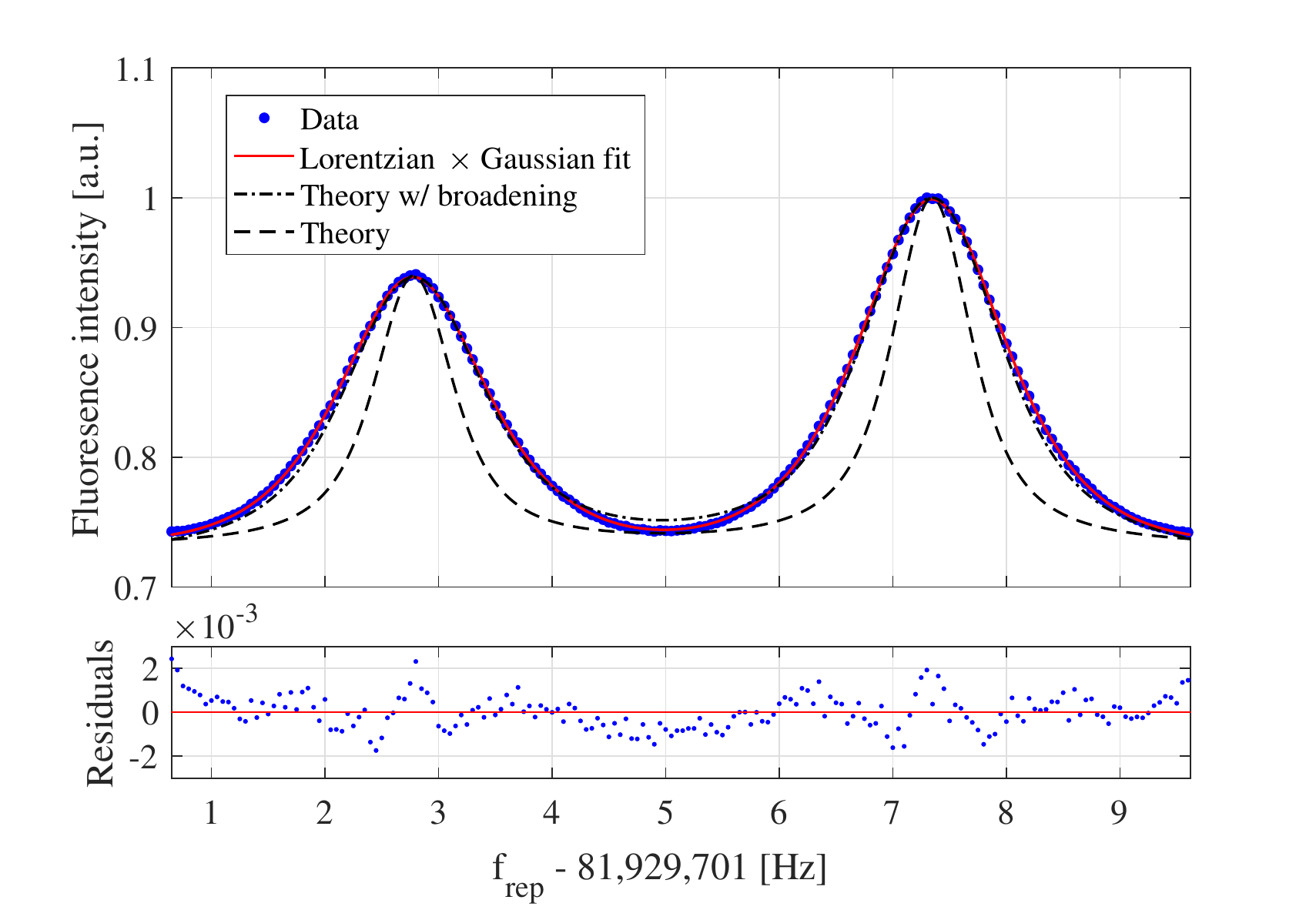}
\caption{(color online) Doppler-free two-photon transition spectrum of $^{133}$Cs $6S_{1/2}$-$8S_{1/2}$ transitions. The left peak corresponds to $F=3\rightarrow F'=3$ transition and the right to $F=4\rightarrow F'=4$. Blue dots are experimental data points (integration of the spatial excitation pattern in Fig.~\ref{fig1}.) and red line is its fit to the product of Lorentzian and Gaussian functions. For this measurement, we used 35~mW of laser power and 10~nm (FWHM) of spectral width and averaged 30 times. The X-axis can be converted to optical frequency domain and 1~Hz in $\Delta f_{\rm rep}$ corresponds to 8.9~MHz in $\Delta f$ [$f = 2f_{\rm ceo}+(m+n)f_{\rm rep},~\delta f = (m+n)\delta f_{\rm rep}, (m+n)\sim 8.9\times 10^6$]. The linewidth is measured to be about 13~MHz (FWHM). }
\label{fig2}
\end{figure}

Figure~\ref{fig2} shows a typical spectrum of the cesium $6S_{1/2}$-$8S_{1/2}$ transitions, where the left peak corresponds to the ${F=3} \rightarrow {F'=3})$ and the right the ${F=4} \rightarrow {F'=4})$. The measured spectrum is compared with a theoretical absorption profile, which is a refined version from Ref.~\cite{BarmesPRL2013}, given by
\begin{widetext}
\begin{equation}
|a_f^{(2)}(f_{\rm rep})|^2 \approx \left\{\sum_{m, n} \left( \frac{|E_{m}(f_{rep}) E_{n}(f_{\rm rep})|^2 }{(f_{\rm ceo}+nf_{\rm rep}-f_{i})^2[(f_{\rm target}-2f_{\rm ceo}-(m+n)f_{\rm rep})^2+1/4\tau_f^2]} \right) \circ G^{D}_{m,n}(f_{\rm rep})\right\}\circ G^{B}(f_{\rm rep})
\end{equation}
\end{widetext}
where $m$ and $n$ are integers and denotes comb index, $\tau_f$ is the lifetime, $\circ$ denotes convolution, $G^{D}_{m,n}(f_{\rm rep})$ = $\exp \left[(\frac{c^{2}m}{2k_{B}T}) (\frac{f_{t}-2f_{\rm ceo}-(m+n)f_{\rm rep}}{(m-n)f_{\rm rep}})^{2}\right] $ is the line-shape from the residual Doppler-shift that comes from the imbalance between the momenta of two modes in each pair, and $G^{B}(f_{\rm rep})$ is the line-shape from the other broadening factors such as the laser linewidth (1.3~MHz), and transit-time broadening (1.6~MHz) in our experimental setup [in total, $G^{B}(f_{\rm rep})$ is the Voigt profile]. The theoretical model contains series of summations so that performing the fitting procedure is rather time-consuming. So, the center frequency of the absorption line was extracted by fitting the data to the product of Lorentzian and Gaussian functions with its coefficients as free parameters. Typical fit and its residuals together with the theoretical lines (including an offset) are shown in the Fig.~\ref{fig2}. The measured linewidth is an order of magnitude broader than the natural linewidth of this transition ($\sim$1~MHz), due to line broadening mechanisms, including the residual Doppler shift, the laser linewidth (1.3~MHz over our integration time), and the transit-time broadening (1.6~MHz). 

Our measurement is summarized in Table~\ref{table1} and compared with previous measurements. Fendel et al. in Ref.~\cite{FendelOL2007} used picosecond lasers frequency-stabilized to a frequency comb laser and Stalnaker et al. in Ref.~\cite{StalnakerPRA2010} used self-reference frequency comb lasers but each counter-propagating beam was color-filtered. Our measurement based on DFCS using  coherent control method agrees well with these results, within the range of uncertainty.  
\begin{table}[h]
\caption{\label{table1}
Summary of the measured frequency of $6S_{1/2}$-$8S_{1/2}$ transition of the $^{133}$Cs. All the frequencies are in MHz unit.}
\begin{tabular*}{\hsize}{@{\extracolsep{\fill}}cccc}
\hline\hline
               & $F=3 \rightarrow F^{\prime}=3$           & $F=4 \rightarrow F^{\prime}=4$\\
\hline
This work   &  729,014,476.90(40)  &  729,006,160.73(33)         \\
Fendel~\cite{FendelOL2007} & 729,014,476.834(15) & 729,006,160.702(15)         \\
Stalnaker~\cite{StalnakerPRA2010} & 729,014,476.65(22)  &   729,006,160.58(22)      \\
\hline\hline
\end{tabular*}
\end{table}

Now we discuss the systematic errors caused by pressure shift, transient-time broadening, Zeeman shift, and AC Stark shift. 

{\it Pressure shift:} Previously measured pressure shifts are $-26$~kHz/mTorr ($-12$~kHz/mTorr) for $F=4$ ($F=3$)~\cite{HagelOC1999}. In our experimental condition, the temperature of the vapor cell was maintained at around 60$^\circ$C and the corresponding vapor pressure was $3\times 10^{-5}$~Torr. This leads to a pressure shift of -780 (-360)~Hz for $F=4$ ($F=3$), which is smaller than our measurement resolution. 

{\it Transit-time broadening:} The beam waist of our laser was around 50~$\mu$m, resulting a transit-time broadening~\cite{Demtroder} of $\delta f_{\rm t}=0.4 v/w=1.6$~MHz, where $w=50$~$\mu$m is the Gaussian beam waist and $v=204$~m/s the most probable speed of atoms. Due to the small beam waist, the transit-time broadening was on a similar order in magnitude of the natural linewidth of $8S$ levels but this does not shift the line centers.

{\it Zeeman shift:} Zeeman shift of this transitions have been measured to be 2~kHz at 10~Gauss~\cite{FendelOL2007}. Our experimental region near the vapor cell was covered with $\mu$-metal so there is negligible magnetic field. So, the Zeeman shift was within our measurement uncertainty.

\begin{figure}[t]
\centering
\includegraphics[width=0.45\textwidth]{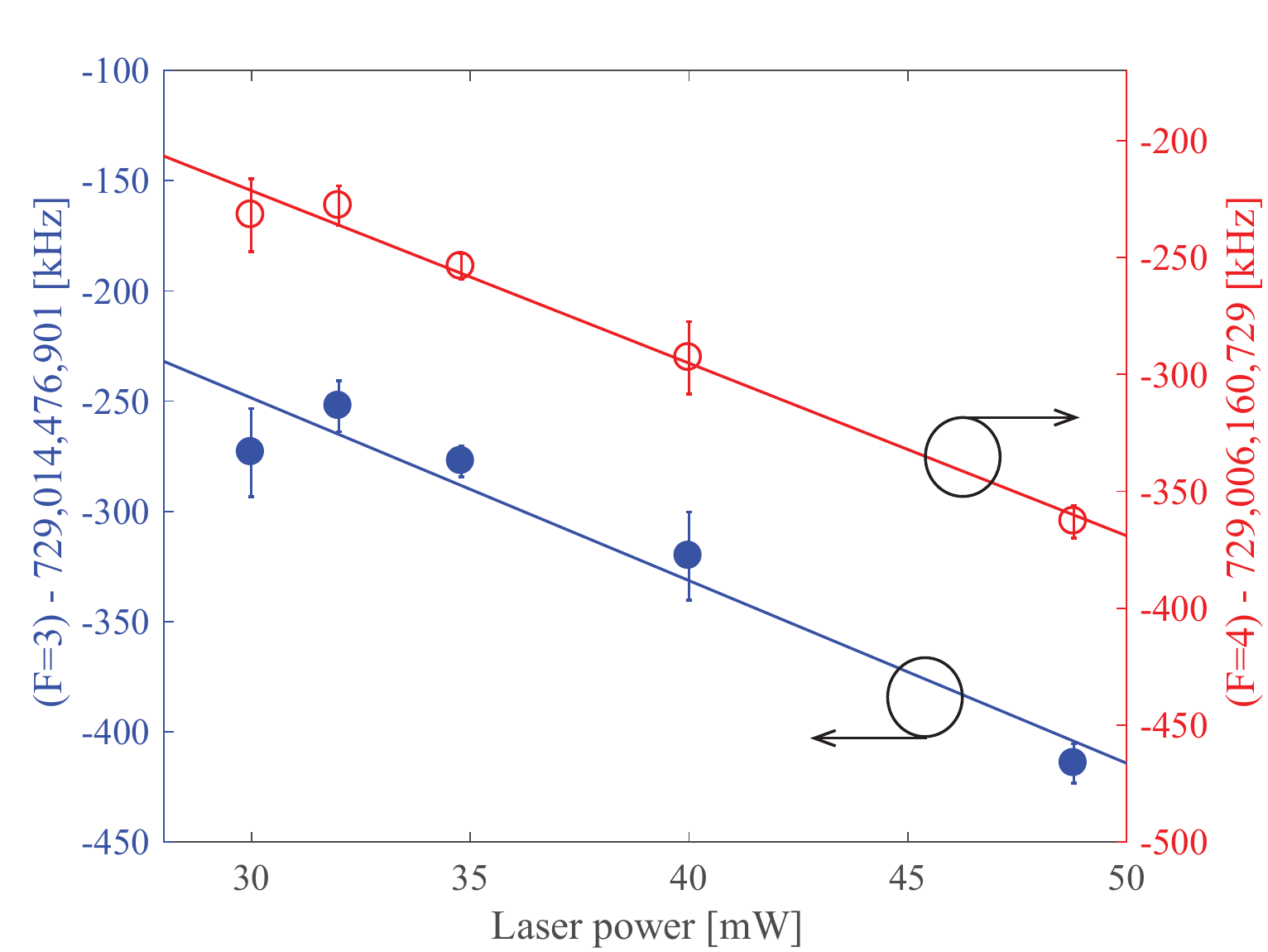}
\caption{(color online) Dependence of the measured center frequency on the laser power. Void (filled) circle corresponds to $F=4$ $(F=3)$ transition frequency. We extrapolate the transition frequencies to zero laser power. The error bar shows a standard error for each fitting.}
\label{fig3}
\end{figure}

{\it AC Stark shift:} AC Stark shifts in these transitions are known to be linear proportional to the average laser intensity (not the laser peak intensity). According to Ref.~\cite{FendelOL2007}, the AC Stark shift is around $-0.21$~Hz/(mW/cm$^2$). In our experiment, the maximum intensity was around ~50~mW for a beam size around 50~$\mu$m. So, the expected shift of the line center is around $-132$~kHz at 50~mW laser power. This is the main, and only considerable, cause of the systematic error to the frequency shift, among the four considered above. Figure~\ref{fig3} shows the Stark shifts measured for various laser intensities (controlled by the pulse shaper). The unshifted frequency is obtained through extrapolating to zero field. The measured slopes for the $F=4$ and $F=3$ transitions were $-7.4 (0.9)$~kHz/mW and $-8.3  (2.0)$~kHz/mW, respectively. We take the shift at typical experimental condition, 34.8~mW, as the systematic error from AC Stark shift of our measurement, and it gives $289$~kHz for $F=4$ and $358$~kHz for $F=3$.

The observed Stark shift is proportional to the average power instead of the peak power that might lead to much larger frequency shift. Tn the context of the time-domain population dynamics~\cite{FelintoPRA2004}, the excited-state population is determined by the phase difference of the sequential excitation probability amplitudes given by the train of pulses, and each excitation is affected by the previous pulses by means of Stark shift. Therefore, the amount of this Stark shift is proportional to the temporal integration of the laser intensity, before the pulse arrives, and it is again proportional to the averaged laser intensity. It is noteworthy, however, that in the strong-field regime experiments, the Stark shift depends on the pulse peak power~\cite{LimSR2014} (enough to induce $Z$ rotations in sub-picoseond time scales).

\section{conclusion}
In summary, we considered direct frequency-comb spectroscopy of $^{133}$Cs $6S_{1/2}$-$8S_{1/2}$ two-photon transition. We utilized the counter-propagating beam geometry of spectrally-encoded ultrafast laser pulses to probe the two-photon transitions of atomic cesium. When being compared with theory and previous measurements, our measured Doppler-free transition profiles are in a good agreement. The absolute frequencies of these transitions are determined with uncertainty below $5\times10^{-10}$ and our error analysis concludes that the main cause of uncertainty comes from laser-induced AC Stark shift.


\end{document}